\documentclass[12pt,a4paper,twocolumn]{narms2}

\usepackage{subfigure} 
\usepackage{psfig}
\usepackage{timesmt}   
\usepackage{graphicx}
\usepackage{amsmath}
\usepackage{amssymb}
\usepackage{chikako}   
\makeatletter
\renewcommand{\section}{\@startsection%
{section}%
{1}%
{0mm}%
{- \baselineskip}%
{0.15\baselineskip}%
{\normalfont\normalsize}}%

\renewcommand{\subsection}{\@startsection
{subsection}%
{2}%
{0mm}%
{-\baselineskip}%
{0.15\baselineskip}%
{\normalfont\normalsize}}%
\makeatother

\newcommand{\el}{{\rm el}}
\newcommand{\dis}{{\rm dis}}

\begin{document}
\title{Adhesive interactions of viscoelastic spheres}
\author{\large {N. V. Brilliantov}\\
{\em Institut f\"ur Physik, Universit\"at Potsdam, Am Neuen Palais 10, 14469 Potsdam, Germany}\\[0.3cm]
\large {T. P\"oschel}\\
{\em Institut f\"ur Biochemie, Charit\'e, Monbijoustra{\ss}e 2, 
  D-10117 Berlin, Germany}
}
\date{\vspace*{-6.9cm} {\begin{flushright}\footnotesize In: Proceedings Powders \& Grains'05, Balkema (Rotterdam, 2005)\vspace*{4cm}\end{flushright}}}

\abstract{We develop an analytical theory of adhesive interaction of viscoelastic spheres in quasistatic approximation. Deformations and deformation rates are assumed to be small, which allows for the application 
of the Hertz contact theory, modified to account for viscoelastic forces. The adhesion interactions are described by the Johnson, Kendall, and Roberts theory. Using the quasistatic approximation we derive the total force between the bodies which is not sufficiently described by the superposition of elastic, viscous and adhesive contributions, but instead an additional cross-term appears, which depends on the elastic, viscous and adhesive parameters of the material. Using the derived theory we estimate the contribution of adhesive forces to the normal coefficient of restitution and derive a criterion for the validity of the viscoelastic collision model.}

\maketitle
\frenchspacing   


\section{INTRODUCTION}

Numerous phenomena observed in granular systems, ranging from sand and powders to granular gases in planetary rings or protoplanetary discs, are direct consequences of the specific particle interactions. Besides elastic forces, common for molecular or atomic materials (solids, liquids, and gases), colliding granular particles exert also dissipative forces.  These   forces acting between contacting grains give rise to unusual properties of granular matter. Hence, the use of an appropriate model of the dissipative interactions is necessary for the  adequate description of granular systems.  In real granular systems the particles may have a complicated non-spherical shape,  differ in size, mass and material properties. In what follows, however, we assume that granular particles are smooth spheres of the same material. We also assume that particles interact exclusively via pairwise mechanical contact.

There are three different types of forces acting between contacting particles: (i) repelling elastic forces, due to the compression of particles, (ii) attractive adhesive forces which appear when particles share a common surface and (iii) dissipative forces, acting against the relative motion of the particles. The dependence on the material parameters and on the quantities which characterize the relative position and motion  of particles is known for all of these forces, e.g. \shortcite{BrilliantovPoeschelWiLey}. The total force acting between the particles is, however, not just the sum of the above three components. Instead, a more careful analysis, sketched below, reveals that the total force contains an additional cross-term which depends on both dissipative and adhesive parameters.  
Using the obtained expression for the total force acting between adhesive viscoelastic spheres, we estimate the effect of the adhesive force on the coefficient of restitution and analyze the range of validity of the frequently used viscoelastic interaction model \shortcite{BrilliantovSpahnHertzschPoeschel:1994,BrilliantovPoeschelOUP}.

\section{FORCES OF GRANULAR PARTICLES}
\begin{figure}[htbp]
  \centerline{\includegraphics[width=5cm,angle=0,clip]{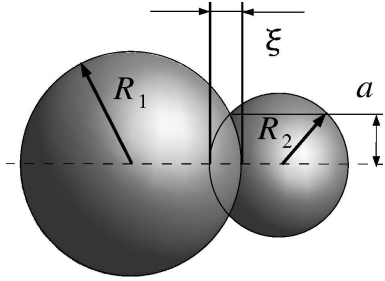}}
  \caption{Collision of spheres, characterized by the time dependent compression $\xi(t) \equiv R_1+R_2 -\left|\vec{r}_1(t)-\vec{r}_2(t)\right|$ and its rate  $\dot{\xi}(t)=v_1(t)-v_2(t)$.}
  \label{fig:psiatcollis}
\end{figure}

{\em Elastic force.} 
Consider a static contact of two spheres of  radii $R_1$ and $R_2$. When the spheres are squeezed, the material in the bulk is deformed (Fig. \ref{fig:psiatcollis}). 
The displacement field $\vec{u}\left(\vec{r}\,\right)$ causes the deformation field
\begin{equation}
\label{u_def}
u_{ij}(\vec{r}\,)=\frac 12 \left( \frac{\partial u_i}{ \partial x_j}  + \frac{\partial u_j}{ \partial x_i} \right) \,,~~~~~i,j =\{x,y,z\}\,.
\end{equation}
In the elastic regime $u_{ij}(\vec{r}\,)$ is proportional to the stress tensor $\sigma^{ij} (\vec{r}\,)$, which gives the $i$-component of the force, acting on a unit surface normal to the direction $j$  \cite{Landau:1965}: 
\begin{equation}
\label{sigma_def_el}
\sigma^{ij} (\vec{r}\,)\!=
\!E_1\!\! \left(\!\!u_{ij}(\vec{r}\,) -\frac13 \delta_{ij} u_{ll}(\vec{r}\,)\!\! \right) \!+E_2 \delta_{ij} u_{ll}(\vec{r}\,)\,.
\end{equation}
Repeated indices imply summation and the elastic coefficients $E_{1}$ and $E_2$ are related to the  Young modulus $Y$ and the Poisson ratio $\nu$ by
\begin{equation}
\label{E1E2}
E_1=\frac{Y}{(1+\nu)}, \quad \qquad E_2= \frac{Y}{3(1-2\nu)} \,.
\end{equation}
The contact problem for elastic spheres has been solved by Hertz \citeyear{Hertz:1882}: the circular contact of radius $a$ gives rise to the elastic force, which read in terms of the compression $\xi\equiv R_1+R_2-\left|\vec{r}_1-\vec{r}_2 \right|$, 
\begin{equation}
\label{eq:a2_Hertz}
a^2=R_{\rm eff} \, \xi \,,
\qquad F_{H} = \frac{\sqrt{R_{\rm eff}}}{D} \xi^{3/2} \,
\end{equation}
where 
\begin{equation}
\label{Spheres}
R_{\rm eff}\equiv\frac{R_1R_2}{R_1+R_2} \,,
\qquad 
D \equiv \frac{3}{2} \frac{(1-\nu^2)}{Y}\, .
\end{equation}
The normal pressure $P_H(x,y)\equiv\sigma^{zz} (x,y,z=0)$, which acts between the compressed bodies in the plane of contact $z=0$ reads \shortcite{Hertz:1882} 
\begin{equation}
\label{Pressure}
P_H(x,y) =\frac{3F_H}{2\pi ab} \sqrt{1-\frac{x^2}{a^2} -\frac{y^2}{b^2}} \,.
\end{equation}

\noindent
{\em Adhesive force.} 
The Hertz theory for the elastic contact of spheres was extended to adhesive contact by Johnson, Kendall, and Roberts (JKR) \citeyear{JKR:1971}. They found that the contact area is enlarged owing to the adhesive force and, thus, introduced an effective Hertz load $F_H$ which would cause this enlarged area. The contact area of radius $a$ corresponds then to the compression $\xi_H$ for the Hertz load $F_H$. In reality, however, this contact radius occurs at the compression $\xi<\xi_H$. The difference between the Hertz compression $\xi_H$ and the actual one, $\xi$, was attributed to the additional stress
\begin{equation}
\label{eq:P_B}
P_B(x,y)\equiv\frac{F_B}{2 \pi a^2} \, \left( 1-\frac{x^2}{a^2}-\frac{y^2}{a^2}\right)^{-1/2}  \, ,
\end{equation}
which is the solution of the classical Boussinesq problem \shortcite{Timoshenko:1970}: This distribution of the normal surface traction gives rise to  a constant displacement over a circular region of an elastic body. The displacement $\xi_B$ corresponding to the contact radius $a$ and the total load $F_B$ are related by
\begin{equation}
\label{eq:xi_B}
\xi_B= \frac{2DF_B}{3a} \,,\qquad D \equiv \frac{3}{2} \frac{(1-\nu^2)}{Y}\,.
\end{equation}
The quantity  $F_B$ stands for the additional attractive adhesive force, which acts against $F_{H}$. In the JKR theory it is expressed by the adhesion coefficient $\gamma$, which is twice the surface free energy per unit area of a solid in vacuum:
\begin{equation}
\label{eq:F_B}
F_B=2\pi a^2 \sqrt{\frac{3 \gamma}{2 \pi D a}} \, .
\end{equation} 
Thus, the total force $F=F_H-F_B$ and the actual compression, $\xi=\xi_H-\xi_B$, read 
\begin{eqnarray}
\label{eq:compres_ad}
&&\xi(a) =\frac{a^2}{R_{\rm eff}} - \sqrt{\frac{8 \pi \gamma D a}{3}} \\
\label{eq:Ftot_via_a}
&&F(a)=\frac{a^3}{DR_{\rm eff}}-\sqrt{\frac{6\pi \gamma}{D}} a^{3/2}\,,
\end{eqnarray}
implying a finite contact radius for $F=0$:
\begin{equation}
\label{eq:a_zero_load}
a_0^3= 6 \pi D \gamma R_{\rm eff }^2\, .
\end{equation} 
For decreasing force ($F<0$) the contact radius decreases until  the minimal value 
\begin{equation}
\label{eq:a_sep}
a_{\rm sep}^3 = \frac14 a_0^3 
\end{equation}
corresponding to maximal (in absolute value) negative force. At this point the particles separate. 
\medskip

\noindent
{\em Viscous force}. 
If the deformation of contacting spheres changes with time, an additional dissipative force arises and the stress tensor contains an additional dissipative component $\sigma^{ij}_\dis$. For small deformation rate $\dot{u}_{ij}(\vec{r}\,)$ it reads \shortcite{Landau:1965}, 
\begin{equation}
\label{Vis_Str}
\sigma^{ij}_\dis (t)=\eta_1 \left[\dot{u}_{ij}(t) -\frac13 \delta_{ij} \dot{u}_{ll}(t) \right] + \eta_2 \delta_{ij} \dot{u}_{ll}(t) \, , 
\end{equation}
where $\eta_1$ and $\eta_2$ are the viscous constants. 

If the impact velocity of the colliding bodies is much smaller than the speed of sound in the particle material and if the characteristic relaxation time of the viscous processes in the bulk of the material is much smaller than the duration of the collision, one can apply the {\em quasistatic}  approximation \shortcite{BrilliantovSpahnHertzschPoeschel:1994}. In this approximation the displacement field $\vec{u}(\vec{r}\,)$ coincides with that for the static case $\vec{u}_\el(\vec{r}\,)$, which is the solution of the corresponding elastic problem. The field $\vec{u}_\el(\vec{r}\,)$ in its turn is completely determined by the time-dependent compression $\xi$, i.e., $\vec{u}_\el=\vec{u}_\el(\vec{r}, \xi)$ \shortcite{BrilliantovSpahnHertzschPoeschel:1994}. Relating $\xi$ and $a$ via Eq. (\ref{eq:compres_ad}) and neglecting a small hysteresis in the very beginning of the contact \shortcite{SmithBozzoloetal:1989}, we obtain for adhesive interaction
\begin{equation}
\label{udot_xidot}
\dot{\vec{u}}(\vec{r},t) \simeq \dot{\xi} \frac{\partial }{\partial \xi} \vec{u}_\el(\vec{r}, \xi) = \dot{a} \frac{\partial }{\partial a} \vec{u}_\el \left( \vec{r}, \xi(a) \right)\, . 
\end{equation}
 The dissipative stress tensor reads, respectively
\begin{equation}
\label{Vis_Str_dotxi}
\sigma^{ij}_\dis = \dot{a} \frac{\partial }{\partial a}  \left[ 
\eta_1 \left(u_{ij}^\el -\frac13 \delta_{ij}u_{ll}^\el\right) +
\eta_2 \delta_{ij} u_{ll}^\el \right] \, .
\end{equation}
From Eqs. \eqref{Vis_Str_dotxi} and \eqref{sigma_def_el} follows the relation between the elastic and dissipative stress tensors in quasistatic approximation \shortcite{Brilliantovetal:2004}, 
\begin{equation}
\label{ST_elas_dis}
\sigma^{ij}_\dis = \dot{a} \frac{\partial }{\partial a} \sigma^{ij}_\el
\left(E_1 \leftrightarrow \eta_1, E_2 \leftrightarrow \eta_2 \right)  \,,
\end{equation}
meaning that the dissipative tensor is obtained from the corresponding elastic tensor by substituting the elastic constants by the viscous constants and applying the operator $\dot{a} \partial / \partial a$. 
In particular the normal component of the stress tensor at the plane $z=0$ for an adhesive contact reads 
\begin{equation}
\label{eq:sumeladh}
\sigma^{zz}_\el(x,y,z=0) = P_H(x,y) -P_B(x,y)  \, ,
\end{equation}
with $P_H(x,y)$ and $P_B(x,y)$ given by Eqs. (\ref{Pressure},\ref{eq:P_B},\ref{eq:F_B}). 

From Eqs. (\ref{ST_elas_dis},\ref{eq:sumeladh}) we find the dissipative stress at the contact plane and, integrating it over the contact area, the dissipative force. Referring for detail to \shortcite{Brilliantovetal:2004}, we present here the final result: 
\begin{eqnarray}
\label{eq:Fdis_adh}
&&F_\dis= \dot{a} \left( A \frac{3a^2}{D R_{\rm eff}} +\frac32 B \sqrt{ \frac{6 \pi \gamma}{D}} \sqrt{a} \right) \\
\label{eq:A}
&&A\equiv \frac13\frac{\left(3\eta_2-\eta_1\right)^2}
{\left(3\eta_2+2\eta_1 \right)}
\left[\frac{\left(1-\nu^2\right)(1-2\nu)}{Y\nu^2}\right]  \\
\label{eq:B_adh}
&&B  \equiv - \frac{(3 \eta_2 -\eta_1) (1+\nu)(1-2 \nu)}{3Y \nu} \,.
\end{eqnarray}
The first term in the rhs of Eq. (\ref{eq:Fdis_adh}) corresponds to the dissipative force in the absence of adhesion. The second term is the corresponding cross-term, which depends on both adhesive and dissipative parameters. 

Remind that the above relations for the dissipative force have been obtained within the simple JKR theory, which is quite accurate in the range of parameters of practical interest \shortcite{AttardParker:1992}.


\section{COEFFICIENT OF RESTITUTION}
An important characteristic of rarefied systems (granular gases \shortcite{BrilliantovPoeschelOUP}) is the coefficient of restitution, which quantifies the loss of mechanical energy for pairwise collisions. It relates the pre-collision relative velocity, $g\equiv v_1-v_2$, to that after the collision, $g^{\prime} = v_1^{\prime} -v_2^{\prime}$:
\begin{equation}
\label{eq:def_Eps}
\varepsilon(g) \equiv - g^{\prime}/g\, .
\end{equation} 
The coefficient of restitution may be evaluated solving the two-body collision problem with given interaction forces, yielding $g^{\prime}$ as a function of $g$.

For small velocities, when the kinetic energy of the relative motion of colliding particles is close to the surface energy of the contact, the adhesive forces may change the coefficient of restitution qualitatively. Indeed, adhesive particles may stay compressed in contact even if the external load vanishes. That is, a tensile force must be applied to separate the particles. The work against this tensile force at the very end of the collision reduces the kinetic energy of the relative motion after the impact, that is, it reduces the coefficient of restitution. For small impact velocity the kinetic energy of the relative motion may be too small to overcome the attractive barrier, i.e., the particles stick together after the collision corresponding to $\varepsilon=0$. 

Consider first a pure viscoelastic collision with the impact velocity $g$ and coefficient of restitution $\varepsilon_{\rm v}$. Using the definition (\ref{eq:def_Eps}), the energy balance reads
\begin{equation}
\label{eq:visco}
\frac12 m^{\rm eff} g^2 
-\frac12 m^{\rm eff} \varepsilon_{\rm v}^2(g) g^2 = W_{\rm dis } \, . 
\end{equation}
where 
\begin{equation}
\label{eq:efmass}
m^{\rm eff}\equiv\frac{m_1m_2}{m_1+m_2} \,.
\end{equation}
The work $W_{\rm dis }$ results from the dissipative force Eq. (\ref{eq:Fdis_adh}) with $\gamma=0$. The corresponding coefficient of restitution $\varepsilon_{\rm v}$ is analytically known \shortcite{SchwagerPoeschel:1998,Ramirez:1999}. Turn now to viscoelastic impacts with adhesion, characterized by the coefficient of restitution $\varepsilon_{\rm ad}$. The energy balance reads
\begin{equation}
\label{eq:viscoad}
\frac12 m^{\rm eff}\! g^2 \!
-\!\frac12 m^{\rm eff} \varepsilon_{\rm ad}^2(g) g^2 \!=\! W_{\rm dis \, A}  \!+\! W_{\rm dis \, B} \!+\! W_{\rm ad} \,,
\end{equation}
where $W_{\rm dis \, A}$ is the work of the dissipative force $F_\dis$ due to the first term in Eq. (\ref{eq:Fdis_adh}) and $W_{\rm dis \, B}$ is the work due to the second term. Finally, $W_{\rm ad}$ is the work due to adhesion, i.e., it results from the adhesive force in the region where the total force Eq. (\ref{eq:Ftot_via_a}) is negative, that is, in the region where the contact radius varies from $a_0$ to $a_{\rm sep}$ \shortcite{BrilliantovPoeschelWiLey}:
\begin{equation}
\label{eq:work_tensil}
-W_{\rm ad} =\int_{\xi(a_0)}^{\xi(a_{\rm sep})} F(\xi) d\xi = \int_{a_0}^{a_{\rm sep}} F(a) \frac{d\xi}{da} da \, .
\end{equation}
Using the approximation $ W_{\rm dis \, A} \approx  W_{\rm dis }$ from Eqs. (\ref{eq:visco},\ref{eq:viscoad})  we obtain the coefficient of restitution for viscoelastic collisions with adhesion:  
\begin{equation}
\label{eq:EpsadEpsvis}
\varepsilon_{\rm ad}^2(g)= \varepsilon_{\rm v}^2(g) -\frac{2( W_{\rm ad}+ W_{\rm dis \, B})}{m^{\rm eff} g^2} \, .  
\end{equation}
$ W_{\rm ad}$ may be found  employing Eq.~\eqref{eq:Ftot_via_a} for the total force $F(a)$, Eq.~\eqref{eq:compres_ad} for the compression, which allows to obtain $d\xi /da$, and Eqs.~(\ref{eq:a_zero_load},\ref{eq:a_sep}) for $a_0$ and $a_{\rm sep}$:
\begin{equation}
\label{eq:W_0}
W_{\rm ad}=q_0 \left(  \pi^5 \gamma^5 D^2 R_{\rm eff}^4\right)^{1/3} \,,
\end{equation}
with the analytically known pure number $q_0 \simeq 0.578$ \shortcite{BrilliantovPoeschelWiLey}.
For small dissipation and small adhesion, $W_{\rm dis \, B}$ may be roughly estimated as
\shortcite{Brilliantovetal:2004}:
\begin{equation}
\label{eq:WadB}
W_{\rm dis \, B}=q_1B \gamma^{1/2} D^{-1/5} g^{8/5} \left(m^{\rm eff} R_{\rm eff}^2
\right)^{3/10} \, ,
\end{equation}
where $q_1 \approx 7.93$.

From Eq.~(\ref{eq:EpsadEpsvis}) we obtain  the condition for the validity of the viscoelastic
collision model:
\begin{equation}
\label{eq:condi_visc} g^2   \gg \frac{2}{m^{\rm eff}} \left(W_{\rm ad} + W_{\rm dis \, B} \right)
\, ,
\end{equation}
which with Eqs.~(\ref{eq:W_0},\ref{eq:WadB}) may be written as
\begin{eqnarray}
\label{eq:gB} &&g^2 \gg g_c^2 = \left(\gamma^5 D^2 R_{\rm eff}^4 \right)^{1/3}/m^{\rm eff}
\\ && B \ll \left( D^2 /\gamma R_{\rm eff}^2 \right)^{1/6}  \left( m^{\rm eff } \right)^{1/2}\, .
\end{eqnarray}
With the above condition for $B$ one can neglect  $W_{\rm dis \, B}$ as compared to $W_{\rm ad}$
for the impact velocity $g$ being of the order of $g_c$ ($g \sim g_c $). Then we obtain
respectively the condition for sticking collision when $\varepsilon_{\rm ad}(g_{\rm st})=0$:
\begin{equation}
\label{eq:condi_stick}
 \frac12 m^{\rm eff}\varepsilon_{\rm v}^2(g_{\rm st})g_{\rm st}^2 = W_{\rm ad}\, .  
\end{equation}
Hence, if $g<g_{\rm st}$ for head-on collisions (vanishing tangential component of the impact velocity), the colliding particles stick together and form a joint particle of mass $m_1+m_2$.

\section{CONCLUSION}  

The collision of adhesive viscoelastic spheres is characterized by (i) the elastic Hertz force, (ii) the dissipative force originating from viscoelastic bulk deformation, and (iii) the adhesive force. We use the continuum model of adhesive contact by Johnson, Kendall, and Roberts \citeyear{JKR:1971} which is adequate in the range of parameters of practical interest. The total force was derived under the approximation of quasistatic deformation, that is, the impact velocity is assumed to be much smaller than the speed of sound in the material and the viscous relaxation time is much smaller than the duration of the collision. This force is not only the superposition of its three components (i-iii), but there appears an additional cross-term, which depends on both viscous and adhesive parameters of the material. 

Using this force we estimated the contribution of adhesive forces to the normal coefficient of restitution as well as the range of validity of the viscoelastic collision model \shortcite{BrilliantovSpahnHertzschPoeschel:1994} and the condition for sticking impact of head-on collisions. 


\end{document}